\documentclass[fleqn,usenatbib]{mnras}
\usepackage{mathptmx}
\usepackage{amssymb}
\usepackage{import}

\usepackage[T1]{fontenc}

\usepackage{graphicx}	
\usepackage{amsmath}	
\usepackage[none]{hyphenat}
\usepackage{bm}
\usepackage{xcolor} 
\usepackage{hyperref}
\usepackage{chngcntr}

\newcommand{\RMS}{$R_{\text{MS}}$} 
\newcommand{\lRMS}{$\log_{10}(R_{\text{MS}})$} 
\newcommand{\lLX}{$\log_{10}(L_{\text{X}})$} 
\newcommand{\lunits}{$L_{2-10}$}
\newcommand{\thh}[1]{$\theta_{#1}$} 
\newcommand{\Thh}{$\boldsymbol{\theta}$} 
\newcommand{\IR}{\textbf{IR}} 
\newcommand{\Xg}{\textbf{X1}} 
\newcommand{\X}{\textbf{X2}} 
\newcommand{\lz}{$0.8~<~z~<~1.2$}
\newcommand{\hz}{$1.0~<~z~<~1.8$}
\newcommand{\savefluxes}{\texttt{savefluxes}}
\newcommand{\fracAGN}{\texttt{frac\_AGN}}
\newcommand{\pdfanalysis}{\texttt{pdf\_analysis}}
\newcommand{\tabA}{Table A1} 

\defcitealias{B19}{B19}
\defcitealias{G20}{G20}
\defcitealias{cigale}{CIGALE}

\newcommand{\G}{\citetalias{G20}}
\newcommand{\B}{\citetalias{B19}}
\newcommand{\BandG}{\B{}~\&~\G{}}
\newcommand{\C}{\citetalias{cigale}}

\title[IR AGN and SFR]{Comparing Infrared \& X-ray-selected Active Galactic Nuclei host galaxy star formation rates to the galaxy main sequence}

\author[D.T. Mason, J.R. Mullaney, L.P. Grimmett]{
Dustin T. Mason\thanks{E-mail: dmason3@sheffield.ac.uk}
James R. Mullaney \thanks{E-mail: j.mullaney@sheffield.ac.uk}
Liam P. Grimmett \thanks{E-mail: lgstatsandresearch@gmail.com}
\\
University of Sheffield, Hicks Building, Sheffield S3 7RH, UK
}

\pubyear{2026}

\begin{document}
\label{firstpage}
\pagerange{\pageref{firstpage}--\pageref{lastpage}}
\maketitle


\begin{abstract}
    \noindent A key method of exploring the relationship between active galactic nuclei (AGN) and their host galaxies is to determine the astrophysical properties of galaxies  that host AGN, and compare them to those that do not.
    Star formation rates are believed to be strongly related to nuclear activity, both indirectly via their common fuel supply, and directly via the energy injected by the AGN either suppressing or enhancing star formation.
    Strong observational evidence of a direct link between nuclear activity and in-situ star formation has proven elusive, likely in part due to the very different timescales of nuclear activity and episodes of star formation. 
    To overcome these challenges, some studies have demonstrated that evidence of the aforementioned link may be found in the distributions of AGN power and star formation rates.    
    While previous attempts to determine these distributions have largely focused on X-ray-selected samples of AGN, in this study we also attempt to determine the star formation rate distributions of infrared-selected AGN. 
    The results vary between our samples, but at low redshift (\lz{}) we find that more powerful X-ray-selected AGN are more likely to occur in galaxies with higher specific star formation rates, consistent with previous studies. At higher redshifts (\hz{}), this trend is reversed with higher power AGN tending to occur in galaxies with lower specific star formation rates. By contrast, infrared-selected galaxies at \hz{} show no evidence of a changing star formation rate distribution with AGN power.
    \end{abstract}



\section{Introduction}\label{sec:intro}

The relationship between active galactic nuclei (AGN) and their host galaxies remains the subject of a great deal of discussion within the astronomy literature \citep[e.g.,][]{ferrarese2000fundamental, hopkins2006fueling, stanley2017mean}. The supermassive black holes (SMBHs) that reside at the centers of most, if not all, galaxies \citep[e.g.,][]{kormendy1995inward} grow over long periods of time by accreting material from the central regions of their host galaxies \citep[e.g.,][]{heckman2004present}. 
This in-falling material releases vast amounts of gravitational potential energy in the form of electromagnetic radiation \citep[e.g.,][]{salpeter1964accretion, 1969Natur.223..690L}, enough in some cases to outshine the host galaxy. 

The inward flow of mass feeding the AGN and the subsequent outward release of energy is often invoked to explain key aspects of galaxy evolution \citep[e.g.,][]{Cattaneo09}, and is required by cosmological-scale simulations to reproduce the properties of today's galaxies \citep[e.g.,][]{croton2006evolution, somerville2008semi, booth2009cosmological, weinberger2018supermassive}. In particular, the decline or cessation of star formation in massive galaxies is not easily reproduced by models of galaxy evolution that do not include energy injection from AGN. As such, the triggering and maintaining of nuclear activity is regarded as having played a key role in shaping today's galaxies.

Despite their importance in our understanding of galaxy evolution, the galactic-scale processes that lead to nuclear activity are not well understood; not least because the central regions of AGN hosts are often obscured, unresolved and/or dominated by the light from the AGN. While there is some evidence of small-scale galactic features such as bars feeding material to the SMBH \citep[e.g.,][]{shimizu2019multiphase}, how these processes themselves form is unclear - as is their ubiquity and relative importance in causing nuclear activity. For example, while mergers may trigger nuclear activity \citep[e.g.,][]{heckman1986galaxy, pierce2023galaxy} by bringing large reserves of cold gas into central galactic regions, not \textit{all} AGN are observed to be in galaxies that show evidence of interactions. As such, galactic-scale internal processes - such as the bulk motion of gas caused by episodes of star formation or galaxy dynamics - likely also play a role in switching AGN on and off \citep{review_BH_bulge_coeval}.

Compounding the difficulties highlighted above regarding studying nuclear activity directly, their innate variability \citep[e.g.,][]{ulrich1997variability} also presents a challenge to investigating the mechanisms that trigger AGN. 
As well as the short term variability that AGN exhibit on human timescales \citep[days to decades; e.g.,][]{sartori2018model}, the fact that most, if not all, currently inactive galaxies host dormant \textit{supermassive} black holes\footnote{Whilst there are numerous models for producing massive black holes in the early Universe \citep{Loeb_collapse, Madau_pop3}, all of them require post-formation growth via accretion to produce the observed SMBH mass function.} implies that they all go through periods of rapid, luminous accretion that are comparatively short-lived on cosmological timescales (i.e. on the order of millions of years).\footnote{Such long-term variability is also commonly referred to as the AGN "duty cycle".}  While long compared to human timescales, this is vastly shorter than those of most other galactic processes, such as episodes of star formation, the formation of internal structures, or even major mergers. This further complicates our attempts to link nuclear activity to galactic-scale triggering mechanisms.

However, insights into AGN triggering mechanisms can still be gained from studying the properties of AGN-hosting galaxies. If, for example, the most powerful AGN were predominantly found in galaxies undergoing major mergers, then this would support the notion that mergers trigger rapid accretion. Along these lines, recent studies that quantify the bulge compactness (central surface mass density) in star forming galaxies have reported that it strongly correlated with AGN accretion rates \citep[e.g.,][]{review_compact_a}, suggesting that the morphology of the inner regions of the host galaxy plays a role in determining the level of AGN activity.

One specific focus of study along these lines has been the consideration of the star-forming properties of AGN host galaxies. Many early studies in this area identify a sample of AGN via some selection criteria (such as emission line ratios, X-ray luminosity, infrared selection as discussed later, etc.), then calculate the average star-forming properties of their hosts. Such studies often find little evidence of a change in the average star-forming properties of AGN-hosting galaxies as a function of AGN power \citep[e.g.,][]{mullaney2012goods, rosario2012mean, stanley15}. However, it was shown in \cite{Hickox14} that this was likely a consequence of the short AGN duty cycle relative to changes in the star-forming properties of their hosts. In that same study, however, they also showed that it may be possible to identify changes in host star-forming properties as a function of AGN power by instead considering the \textit{distributions} of star formation rates (SFRs)\ rather than average star formation rates alone. 

Following the aforementioned \cite{Hickox14} study, \cite{B19} (hereafter \B{}) used a maximum likelihood approach to determine the distributions of the specific star formation rates (sSFRs) - i.e. the total star formation rate of a galaxy in solar masses per year, normalised by galaxy mass - of AGN-hosting galaxies separated into two bins according to X-ray luminosity. 
As predicted by the \cite{Hickox14} study, \B{} found marginal (sub-$3\sigma$) evidence that the sSFR distribution of higher power AGNs is narrower and shifted to higher values than lower power AGNs. 
Later, \cite{G20}, hereafter \G{}, analysed the same sample of AGN with a binning-free hierarchical Bayesian model and obtained results that were consistent with \B{} - i.e. that sSFR distribution of AGN hosts relative to typical non-AGN host galaxies (measured using the metric \RMS{}; see \autoref{sec:RMS} for more details on \RMS{} and the definition of ``typical'' galaxies) is dependent on the power of the AGN - but this time with a greater degree of significance. They also showed in figure 4 of \G{} that this connection between \RMS{} and AGN power is able to reproduce the flat relationship between average SFR (non-normalised by galaxy mass) and X-ray luminosity found by \cite{stanley15}.

The overall conclusion reached by \BandG{} is that nuclear activity and star formation are linked, such that AGNs with higher X-ray luminosities tend to reside in galaxies with higher sSFRs. This is consistent with the findings of \cite{aird2018x, aird2019x} who find that galaxies with increased levels of star-forming activity (i.e., those lying above the main sequence) have a higher chance of hosting a luminous AGN. To date most work on determining the distributions of sSFRs of AGNs have focused on X-ray-selected AGN, largely due to the ease of their selection. However, X-ray selection is biased against the heavily obscured AGN that are now known to represent a significant proportion of the AGN population and, subsequently, the accretion history of the Universe \citep[e.g.,][]{aird2019x}.

Over the past few decades, it has been demonstrated that infrared (hereafter referred to as IR) wavelengths provide a viable means of selecting heavily obscured AGN, although no single wavelength band can select all AGN \citep[e.g.,][]{lyu2022agn}. At IR wavelengths, AGN manifest through power-law like spectral energy distributions (SEDs). Should this dominate over the host galaxy at these wavelengths, then it can be used as a means of identifying AGN. With IR wavelengths being less susceptible to obscuration than X-rays, IR selection can identify heavily obscured AGN that are missed at most other wavelengths. Whilst IR selection may be biased in favour of more luminous AGN, and not all IR-selected AGN are heavily obscured \citep[e.g.,][]{review_selection}, the fact that IR-selection does capture AGN that are not identified via other means means that their study is worthwhile if we are to attempt get the fullest understanding of the relationship between nuclear activity and the properties of their host galaxies. With this in mind, we build upon the work of \BandG{} by investigating how the distributions of the star-forming properties of IR-selected AGN change as a function of AGN power and comparing to X-ray-selected AGN. Differences between samples defined by these selection methods may indicate that IR selection is in fact sampling a distinct population of AGN, and if so, shed light on the relationship between AGN and their hosts.

This paper is laid out as follows: our data sources and sample selection are described in \autoref{sec:data}, followed by our analysis methodology in \autoref{sec:method}. We show and discuss our results in \autoref{sec:results} \& \autoref{sec:discussion} respectively, and summarise our key findings in \autoref{sec:conclusions}.

\section{Data}
\label{sec:data}

For statistical studies such as that carried out here, we wish to avoid biases associated with pre-selection as much as possible. For this reason, we turn to so-called ``blank field'' surveys, targeting areas of the sky with minimal contamination from stars to allow for extragalactic observations. Following \BandG{}, we use the COSMOS field \citep{scoville2007cosmic} due to its good compromise between depth and area, as well as its broad multi-band coverage spanning most of the observable electromagnetic spectrum.

Both \BandG{} used X-ray detected AGN to compare AGN with non-AGN galaxies by selecting all sources with 2-10~keV luminosity, $L_{2-10\text{keV}}$~erg\,s$^{-1}$, greater than $10^{42}$~erg\,s$^{-1}$; this effectively selects AGN as other processes struggle to emit above this luminosity threshold. We also refer to log$_{10}(L_{2-10\text{keV}}$~erg\,s$^{-1}$) when log-space is appropriate, or $L_{\text{X}}$ and \lLX{} respectively for short.

While the primary focus of this study is IR-selected AGN, we also define X-ray-selected samples to compare against. For these, we repeat the same selection process using absorption-corrected X-ray luminosities from the Chandra COSMOS-Legacy Survey catalogue presented in \cite{CCOSMOS}. As in \G{}, we do not include sources with upper limits on X-ray luminosity, nor account for uncertainties in X-ray luminosities arising from redshift uncertainties, although we do account for uncertainties in the X-ray luminosities arising from the uncertainty in the measured X-ray flux.

While our selection process is the same as \B{} \& \G{}, the photometric data we used for SED fitting (the motivation and process of which is described in \autoref{sec:mass}) and IR selection is from a more recent catalogue: the COSMOS2020 catalogue (with \texttt{The Farmer} source extraction) from \cite{weaver}. As such, while very similar, our X-ray-selected samples of AGN are not identical to those used in previous studies.

We apply the criteria defined in \cite{Donley} on the Spitzer IRAC fluxes in the \cite{weaver} catalogue to obtain a sample of IR-selected AGN. More details and possible biases caused by this selection process are explored in \autoref{sec:data:IR}. AGN in the IR-selected sample do not necessarily have X-ray detections due to obscuration of the AGN by gas and dust in the host galaxy, making the IR a powerful probe of these obscured galaxies.

Since the aim of this study is to investigate the distribution of star-forming properties of AGN-hosting galaxies, we also require measured SFRs of the host galaxies of the AGN in our X-ray and IR-selected samples. 
For these, we take SFRs from the Super-Deblended catalogue presented in \cite{SD}. The SFRs included in this catalogue are derived from far-IR \textit{Herschel} \citep{pilbratt2010herschel, lutz2011pacs} observations of the COSMOS field \citep{scoville2007cosmic}, with far-IR photometric fluxes extracted at the positions of sources detected in MIPS-24~{${\rm \mu m}$} \citep{rieke2004multiband, le2009deep}, radio \citep[VLA,][]{thompson1980very, condon1998nrao, schinnerer2010vla, smolvcic2017vla}, 850 $\mu$m images \citep{2017ApJ...837..139C, 2017MNRAS.465.1789G}, 1.1 mm \citep{2011MNRAS.415.3831A}, 1.2 mm \citep{2007ApJS..172..132B}, and optical observations of the field from \cite{laigle2016cosmos2015}. Potential contamination from the AGN is accounted for by fitting the IR SED. This yields more robust SFRs in the presence of AGN than those from the Weaver photometric catalogue, which adopts a more simplistic -- but much faster -- AGN model in its SED fitting by using simplified SED templates and a less flexible framework. 
The relative depth of the \textit{Herschel} data compared to that of the MIPS-24~{${\rm \mu m}$}/radio/optical data from which the prior positions are taken results in a large proportion of upper limits on the SFRs, as sources are not detected by \textit{Herschel}. As we describe in \autoref{sec:hypersampler}, these upper limits are statistically accounted for in our analysis.

\subsection{Samples}\label{sec:samples}

Our samples of AGN are made using two different selection methods: X-ray and IR. These samples are then split into redshift bins for analysis; one redshift bin is chosen to be consistent with \B{} \& \G{} (\lz) for direct comparison. However, because we use more recent photometry and SED fitting code than \G{}, we primarily use this sample to validate that our methods produce consistent results with this earlier study before employing them to analyse the IR sample and other redshift ranges.

Reflecting the focus of this study on comparing X-ray and IR-selected AGN, we then analyse AGN in a second redshift bin that spans \hz{}. This has the benefit of containing substantially more IR-selected AGN than the \lz{} bin.

In summary, we have three samples of AGN:

\begin{enumerate}
    \item \Xg: X-ray-selected AGN within \lz; 323 AGN, 240 upper limits ($\sim$74 per cent) on SFR;
    \item \X{}: X-ray-selected AGN within \hz; 498 AGN, 317 upper limits ($\sim$64 per cent) on SFR;
    \item \IR: IR-selected AGN within \hz; 246 AGN, 129 upper limits ($\sim$52 per cent) on SFR.
  \end{enumerate}

As we regularly refer to, and compare between, these samples, we include shorthand names (in bold). We analyse these samples as described in \autoref{sec:method}, and compare the results from each in \autoref{sec:results}.

\subsection{\RMS{}}\label{sec:RMS}

It is known that the SFRs of typical star-forming galaxies are strongly linked to redshift and stellar mass; \cite{MS} define a galaxy ``main sequence'' (MS) for non-AGN, non-starburst, star forming galaxies that shows increasing SFR with both stellar mass and redshift. As such, when comparing AGN hosts across a range of redshifts and stellar masses, it is important to account for these effects. To achieve this, we use the same \RMS{} statistic employed by \B{} \& \G{}, first defined in \cite{elbaz2011goods} as the ``starburstiness'' of a galaxy. We calculate the \RMS{} of an AGN host galaxy by dividing its SFR (${\rm SFR_{AGN}}$) by the SFR it would have were it to sit on the locus of the MS (i.e., same mass and redshift as a MS galaxy):

\begin{equation}\label{eq:RMS_simple}
    R_{\text{MS}} = \frac{ \text{SFR}_{\text{AGN}} }{ \text{SFR}_{\text{MS}} }.
\end{equation}

\noindent By using this equation, and equation 9 from \cite{MS} to estimate the SFR of a MS galaxy (SFR$_{\text{MS}}$), we calculate the \RMS{} values for all the AGN in our samples. sSFR is not required, as the AGN and model MS galaxy have the same mass.

It has been found that the galaxy MS and potentially AGN host galaxies (discussed in \autoref{sec:hypersampler}) follow a log-normal \RMS{} distribution; for this reason, we often refer to \lRMS{} where appropriate.

\subsection{IR selection}\label{sec:data:IR}

As described at the start of this section, IR selection is performed using equations 1 and 2 from \cite{Donley}, which makes use of the \textit{Spitzer}/IRAC IR bands centred at 3.6, 4.5, 5.8 \& 8.0~$\mu$m. This defines selection criteria in colour space, and ensures that the IR spectrum of the candidate rises with increasing wavelength. This rise is typically attributed to the presence of an AGN \citep[e.g.,][]{mullaney2011defining}. IR selection is well-suited to identifying heavily obscured AGN and \cite{review_selection} report that it is robust to contamination from non-AGN. However, that same study also maintains that IR-selection of this kind will preferentially select more powerful AGN. Along similar lines, our IR selection may result in the exclusion of those cases in which the host galaxy dominates over the AGN at infrared wavelengths, such as those in the the most massive or most rapidly star-forming galaxies.

In order to investigate the effects of such biases on our analyses, we created a synthetic sample of $10^5$ AGN-plus-galaxy SEDs using \C{} (Code Investigating GALaxy Emission, \citealt{cigale}) in simulation (\savefluxes{}) mode. To generate the SEDs, we uniformly randomly select parameters that control the star formation history (which, in turn, implies the current SFR) and the contribution of the AGN relative to the total luminosity of the system (i.e., stellar component plus AGN); the parameter ranges are provided in \autoref{tab:cigale} \& \tabA{}. The masses generated by \C{} are normalised, so we create model masses based on the properties of our observed samples by assuming a lognormal distribution centred on $10^{9}~{\rm M_{\odot}}$ with a $\sigma$ of 0.5 in log-space.

\begin{table}
    \caption{
    \C{} modules used in our analysis for all samples, and the range of parameters used to generate our mock sample. For the analysis of our real AGN samples, \C{} is used in \pdfanalysis{} mode, and the possible parameters used are different to those displayed in this table due to computational constraints (parameters shown in \tabA{}). The mock sample generated from the possible parameters listed here was designed to evenly sample the relevant parameter space, and so differ from the possible parameters used in the analysis. For the mock sample, \C{} was used in \savefluxes{} (simulation) mode, and in this table ``all possible values'' indicates when all possible values supported by the relevant \C{} module (see included references) are used, otherwise a range is given. Parameter values are uniformly sampled from these possible values/ranges of values. Also used is the \C{}-standard redshifting module (which accounts for redshifting of spectral features and absorption by the intergalactic medium) Of note is the \fracAGN{} parameter, which represents the AGN-to-galaxy bolometric luminosity ratio.}
    \centering
    \begin{tabular}{ll}
    \hline
    Module \& parameters &  \\ \hline

    \textbf{Star formation history} & \cite{cigale}                     \\
    $\tau_{main}$                       & $100 - 1\times 10^{10}$ Myr   \\
    age$_{main}$                        & $100 - 12000$ Myr             \\
    $\tau_{burst}$                      & $1 - 1000$ Myr                \\
    age$_{burst}$                       & $1 - 1000$ Myr                \\
    $f_{burst}$                         & $0.001 - 0.99$                \\ \hline

    \textbf{Stellar population synthesis}   & \cite{bruzual2003stellar}                    \\
    IMF                                     & 1 (\citealt{chabrier2003galactic})           \\ 
    Metallicity                             & all possible values                          \\ 
    Separation age                          & $0 - 1000$ Myr                               \\ \hline

    \textbf{Dust attenuation} & \cite{calzetti2001dust}                    \\
    E(B-V)* young stellar population    & $0.01 - 1.4$              \\
    E(B-V)* old stellar population      & $0.01 - 0.99$             \\
    UV bump amplitude                   & $0.01 - 3$                \\
    Power-law slope                      & $0 - 3$                   \\ \hline

    \textbf{Dust emission} & \cite{dale2014}                    \\
    $\alpha$ slope                      & all possible values              \\ \hline 

    \textbf{AGN} & \cite{fritz2006}                           \\
    Torus max-min radius ratio          & all possible values \\
    $\tau$ (9.7$\mu$m optical depth)    & all possible values \\
    $\beta$                             & all possible values \\
    $\gamma$                            & all possible values \\
    opening angle                       & [60, 100, 140]      \\  
    $\psi$ (viewing angle)              & all possible values \\
    \fracAGN{}                     & $0.0 - 1.0$         \\ \hline

    \textbf{Redshift}          &                \\
    $z$                        & $0.8 - 1.8$    \\ \hline

    \end{tabular}
    \label{tab:cigale}
\end{table}

By the nature of how they were generated (i.e., via the uniform random sampling of \C{} parameter space), the synthetic galaxies have a \lRMS{} distribution that is not representative of reality. To produce a sample with a more realistic \lRMS{} distribution, we performed a weighted random subsampling of our initial sample, with the weighting of a given galaxy proportional to the likelihood that it would belong to an arbitrary, non-\lLX{} dependant \lRMS{} distribution. 

\begin{figure}
    \includegraphics[width=\columnwidth]{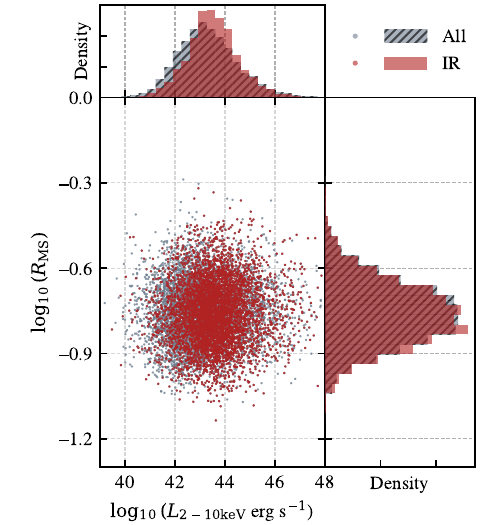} 
    \caption{
        \lRMS{} against X-ray luminosity for a subsample of the full synthetic sample described in \autoref{tab:cigale}, constructed using weighted random sampling to produce a log-normal \lRMS{} distribution with no dependency on AGN power, represented by its X-ray luminosity \lLX{}.
        All refers to the entire sample of AGN, whereas IR refers to the subset of that sample that are selected by the Donley criteria \citep{Donley}.
        There is a slight bias towards more powerful AGN (with higher X-ray luminosities) in IR-selected AGN, but no evidence of a bias in \lRMS{}, indicating the Donley selection criteria does not impose any biases in the \lRMS{} distribution when selecting AGN using IR.
    }
    \label{fig:IRtest}
\end{figure}

Next, we passed our subsample of simulated SEDs (i.e., those that have been forced by subsampling to have a log-normal \RMS{} distribution) through the Donley IR selection criteria to flag those that would be identified as IR-selected AGN. To determine whether IR selection leads to a biased sampling of the underlying \lRMS{} distribution, we plot \lRMS{} against AGN X-ray luminosity in \autoref{fig:IRtest} for the whole subsample of synthetic SEDs, while highlighting those that would be IR-selected. \autoref{fig:IRtest} shows that while IR selection preferentially selects AGN with slightly higher X-ray luminosities \cite[in agreement with][]{review_selection}, the \RMS{} distribution of IR-selected AGN is the same as the underlying population, thereby indicating that IR selection does not result in a biased sampling of \RMS{}.

\begin{figure}
    \includegraphics[width=\columnwidth]{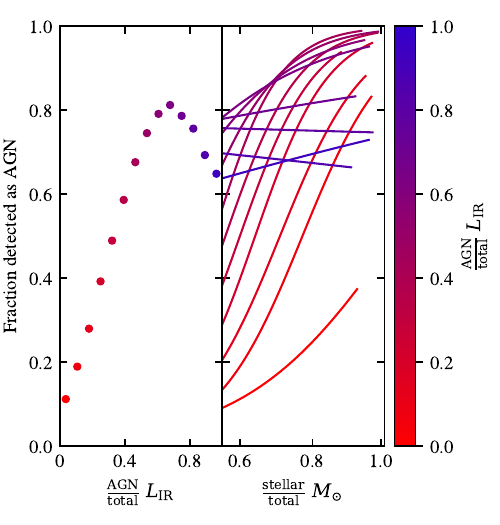} 
    \caption{
        AGN detection fraction according to the Donley criteria \citep{Donley} for simulated AGN \& non-AGN host galaxy SEDs generated using \C{} with a range of astrophysical properties as shown in \autoref{tab:cigale}.
        The y-axis on both subplots show the fraction of simulated galaxies that are detected as AGN by the Donley criteria; the left subplot shows this against the relative contribution of the AGN to the total IR luminosity, while the right subplot shows this against the fraction of the galaxy mass that is held in stars (total galaxy mass = stellar + gas + dust).
        These show the Donley criteria is biased towards detecting AGN that contribute a larger proportion of the total galaxy luminosity in the IR range, as would be expected, but peaking at around 80 per cent before dropping again. 
        The right plot shows that there is a progressive change in the detection rate with both the stellar mass fraction and AGN IR luminosity fraction shown in the left plot and colourbar.
        The fraction of sources selected as IR AGN goes down at very high AGN/total $L_{\rm IR}$ ratio because there are a number of high AGN/total $L_{\rm IR}$ SEDs that sit substantially above the Donley selection criteria, as shown in \autoref{fig:cc_plot}. 
    }
    \label{fig:IR_regression}
\end{figure}

\begin{figure}
    \includegraphics[width=\columnwidth]{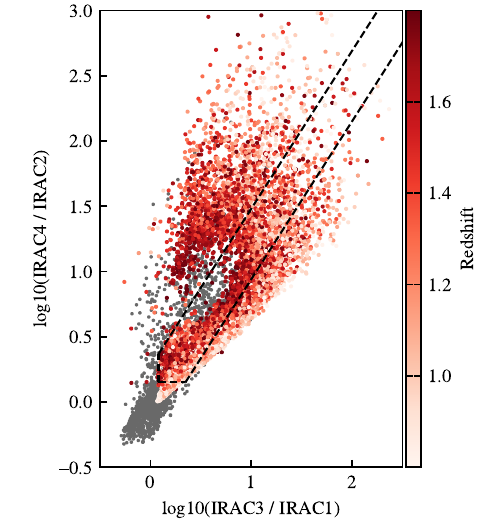}
    \caption{\textit{Spitzer}-IRAC colour-colour plot for the sample of simulated AGN \& non-AGN host galaxy SEDs generated using \C{} with a range of astrophysical properties as shown in \autoref{tab:cigale}. The grey points represent the full sample of SEDs ranging from zero to 80 per cent AGN/total $L_{\rm IR}$ ratios, whereas the red shaded points represent SEDs with $>80$ per cent AGN/total $L_{\rm IR}$ ratios at various redshifts (indicated via the colour bar). The grey points have been downsampled by a factor of twenty for visual clarity. A substantial proportion of simulated SEDs with high AGN fractions lie outside the Donley selection criterion, indicated by the dashed black box. Given that this plot only includes data for simulated SEDs, it should not be interpreted as indicating how common such extreme outlying systems are in the real Universe; rather, this plot is merely used to explain the drop in the proportion of AGN-selected galaxies toward very high AGN fractions shown in \autoref{fig:IR_regression}.}
    \label{fig:cc_plot}
\end{figure}

We show in \autoref{fig:IR_regression} the fractions of our full synthetic sample that are IR-selected as a function of: (a) the contributions of the AGN relative to the total IR luminosity of the system in the left subplot, and (b) stellar-to-total (i.e. stellar, gas, dust) mass ratio in the right subplot. As expected, the IR-selected fraction is most sensitive to the relative contribution of the AGN at IR wavelengths. However, we find that once the AGN contribution reaches $\sim 50$ per cent or more, then $\gtrsim70$ per cent are IR-selected. At these high AGN fractions, we find that the proportion of SEDs that are IR-selected is not strongly affected by the stellar-to-total mass ratio. In \autoref{fig:IR_regression}, the detection fraction goes down at very high AGN/total $L_{\rm IR}$ \citep[i.e., AGN fraction, or the \fracAGN{} parameter in][]{fritz2006} because there are a number of high AGN fraction SEDs that sit substantially above the Donley selection criteria, as shown in \autoref{fig:cc_plot}. 

\section{Method}\label{sec:method}

The focus of this study is on the relationship between AGN and the star forming properties of their hosts. Rather than using SFRs, however, we use \RMS{} to allow direct comparison between galaxies of different masses and at different redshifts. The precise definition and motivation of \RMS{} is described in \autoref{sec:RMS}. Since calculating \RMS{} requires knowledge of stellar masses, we describe how we calculated the latter in \autoref{sec:mass}. In \autoref{sec:LIR} we describe how we find directly comparable AGN powers for all of our samples. Finally, we describe how we determine relationships between the \RMS{} distribution and AGN power in \autoref{sec:hypersampler}.

\subsection{Determining stellar masses}\label{sec:mass}

As previously mentioned, in order to calculate \RMS{} both the SFRs and the stellar masses of each of the AGN host galaxies in our samples must be known. Methods of standard catalogue-wide estimation of astrophysical parameters, as used in \cite{weaver}, are designed to fit catalogues containing large numbers of galaxies, only a small fraction of which host AGN. 
They therefore typically use simplified templates or models of AGN emission, and so the resulting properties are not suitable for studies that focus specifically on AGN hosts as described in \autoref{sec:data}.

The two properties required here, stellar mass and SFR, can both be heavily contaminated by emission from the AGN. We therefore require a more robust method to estimate the mass and SFR of the AGN host galaxies. SFRs are taken directly from \cite{SD}, and are calculated by estimating the relative contributions from star formation and AGN to the IR flux using AGN emission templates from \cite{mullaney2011defining}. 

For stellar mass estimates we again use \C{}, now in analysis mode (\texttt{pdf\_analysis}), which generates a large number of models and compares them to photometric data in order to estimate the SED of the galaxy before using reduced $\chi^{2}$ (accounting for photometric upper limits) to determine which model best fits the observed data. 
These models are drawn from a set of emission and absorption modules, e.g. the Fritz AGN emission model, \cite{fritz2006}; Dale dust emission/absorption model, \cite{dale2014}; all modules are listed in \autoref{tab:cigale}. 
The contribution from each module is in turn controlled by a set of parameters; the range and spacing of the parameter values are defined manually, and thus the regions of parameter space explored by \C{} is set by the user. 

The modules and parameters used for simulating the mock sample are shown in \autoref{tab:cigale} and those used for the main sample analysis are shown in \tabA{}.

After scaling by redshift (and shifting) to account for dimming, the best \C{} model returns a set of astrophysical parameters of the galaxy, drawn from the requisite model parameters and scaling.

Across our sample, the resulting stellar masses are roughly consistent - with a large spread but little to no offset - to those obtained by \cite{weaver} using the photometric fitting codes they apply to the entirety of their catalogue: LePHARE \citep{arnouts1999measuring, ilbert2006accurate, 2011ascl.soft08009A} and EAZY (\citealt{brammer2008eazy}).
Whilst LePHARE and EAZY are appropriate for fitting galaxies, they are not as subtle or configurable in their approach to AGN as \C{}. By contrast, \C{} offers a more robust fit for AGN, which results in the scatter we see when comparing its output to LePHARE or EAZY. The lack of a systematic offset implies that all methods are in broad agreement with each other (in terms of stellar mass estimation), but \C{} should offer more robust results for individual galaxies whilst the simpler methods only recreate the mass distribution.

\subsection{Estimating AGN power}\label{sec:LIR}

To investigate the possible relationship between AGN power and the host galaxy SFR, we must find suitable measures of the AGN power. For those AGN with X-ray detections, we can use the absorption-corrected \lunits{}~keV luminosity from \cite{CCOSMOS}, described in \autoref{sec:data}. \lunits{}~keV luminosity is a suitable proxy for AGN power \citep{lx_power}, but X-ray detections are limited for the AGN in our IR-selected sample despite the full X-ray coverage in the COSMOS field. Indeed, many (154 out of 246, or $\sim$63 per cent) of our IR-selected galaxies are not detected in X-rays indicating greater chances of selecting obscured AGN when using IR-selection methods.

In order to estimate the X-ray luminosity of these AGN, we use the relationship between 12$\mu$m and X-ray luminosity shown in equation 2 of \cite{gandhi2009}.
This allows effective estimation of the X-ray luminosity, and therefore AGN power, of the IR samples. We take the rest-frame 12$\mu$m luminosity from the AGN component of our \C{} SED fits described in \autoref{sec:mass}, propagating the relative uncertainty on the 12$\mu$m luminosity to the derived X-ray luminosity.

\subsection{Hypersampler: Binning-free analysis}\label{sec:hypersampler}

Following \BandG{}, the purpose of this study is to identify any dependency between the parameters of the \RMS{} distribution of the AGN in our samples and an independent variable. Following \B{} \& \G{}, we model the \RMS{} distribution as a log-normal for ease of comparison against the log-normal \lRMS{} distribution of MS galaxies similarly to \cite{sargent2012contribution}. Within the frequentist regime, this can be thought of as a probability distribution. In the case of sources with statistically significant \RMS{} measurements, this corresponds to the following probably density function (PDF): 
\begin{equation}
f\left(\log_{10}(R_{MS})|\mu,\sigma\right) = (2\pi\sigma^2)^{-1/2}{\rm exp}\left(-\frac{\left(\log_{10}\left(R_{\rm MS}\right)-\mu\right)^2}{2\sigma^2}\right),
\end{equation}

\noindent
where $\mu$ and $\sigma$ are the mean and variance of the log-normal distribution, and the logarithms are base 10. By contrast, in the case of sources that only have upper limits on their \RMS{} measurements (principally due to them being undetected at far-infrared wavelengths), then the PDF is replaced by the cumulative distribution function (CDF), which is the integral of the above PDF:
\begin{align}
    F\left(\log_{10}(R_{MS})|\mu,\sigma\right) &= \int_{-\infty}^{\log_{10}(R_{\rm MS})}f\left(\log_{10}(R_{\rm MS})|\mu,\sigma\right)d\log_{10}(R_{\rm MS}) \nonumber \\ 
    &=\frac{1}{2}\left(1 + {\rm erf}\left(\frac{\log_{10}(R_{\rm MS})-\mu}{\sigma\sqrt{2}}\right)\right),
\end{align}
where ${\rm erf()}$ refers to the Gaussian error function. The likelihood function for the full sample of $n$ AGN, of which $m$ have statistically significant \RMS{} measurements, is the product of the above PDFs and CDFs:
\begin{align}
L(\log_{10}(R_{{\rm MS}})|\mu,\sigma) = &\prod_{i=1}^m f\left(\log_{10}(R_{{\rm MS},i})|\mu,\sigma\right) \nonumber \\
& \times \prod_{i=m+1}^n F\left(\log_{10}(R_{{\rm MS},i})|\mu,\sigma\right),
\end{align}\label{eq:likelihood}

\noindent
where $R_{{\rm MS},i}$ is the measured \RMS{} value or upper limit for the $i^{\rm th}$ source in the sample. This treatment of detections and upper limits within the likelihood function allows us to properly account for non-detections (upper limits). Since a large proportion of our \RMS{} values are upper limits, this allows us to incorporate far larger datasets and therefore acquire more robust statistics; while an upper limit is not as ``valuable'' as a detection, it can still contribute to constraining the \lRMS{} distribution.

With the likelihood function defined, we next attempt to ascertain whether the parameters of the above distribution change with AGN power. This could mean that either $\mu$ or $\sigma$ change with AGN power, or that both do. Following \G{}, we use simple first-order polynomials to describe a possible relationship between $\mu$ and AGN power, and $\sigma$ and AGN power:

\begin{align}\label{eq:mu_sigma}
    \mu_{i}(X_{i}) &= \theta_{0} + \theta_{1}  X_{i}; \nonumber \\
    \sigma_{i}(X_{i}) &= \text{e} ^{\theta_{2} + \theta_{3} X_{i}}
    ,
\end{align}

\noindent where $X_{i}$ represents a measure of the power of the $i^{\rm th}$ AGN in our sample. The use of the exponent in the parameterisation of $\sigma$ is simply to ensure that $\sigma$ cannot become negative. In this study we use $X=\log_{10}\left(L_{\rm X}/10^{40}~{\rm ergs~s^{-1}} \right)$, where $L_{\rm X}$ is the $2-10~{\rm keV}$ luminosity of the AGN, either determined directly from X-ray observations in the case of our X-ray-selected samples, or converted from the $12~{\rm \mu m}$ luminosities following \cite{gandhi2009}. Like \G{}, we account for the uncertainty on these X-ray luminosity measurements but exclude galaxies with X-ray upper limits, i.e. non-detections.

From \autoref{eq:mu_sigma}, it is clear that $\theta_1$ and $\theta_3$ control how strongly $\mu$ and $\sigma$ - and thus, in turn, the \lRMS{} distribution itself - vary with increasing AGN power. If $\theta_1=0$, then $\mu$ has no dependency on AGN power. Similarly, if $\theta_3=0$, $\sigma$ has no dependency on AGN power. For convenience, we define  \Thh{}$~=~\{\theta_{0}, \theta_{1}, \theta_{2}, \theta_{3}\}$.

Finding the values of \Thh{}, together with their uncertainties, is possible using a standard implementation of MCMC sampling. However, in order to determine whether there is significant evidence that the \lRMS{} distribution changes with AGN power, we have developed a sampling code called Hypersampler, which we describe next (see also \G{}).

\begin{figure*}
    \includegraphics[width=1.95\columnwidth]{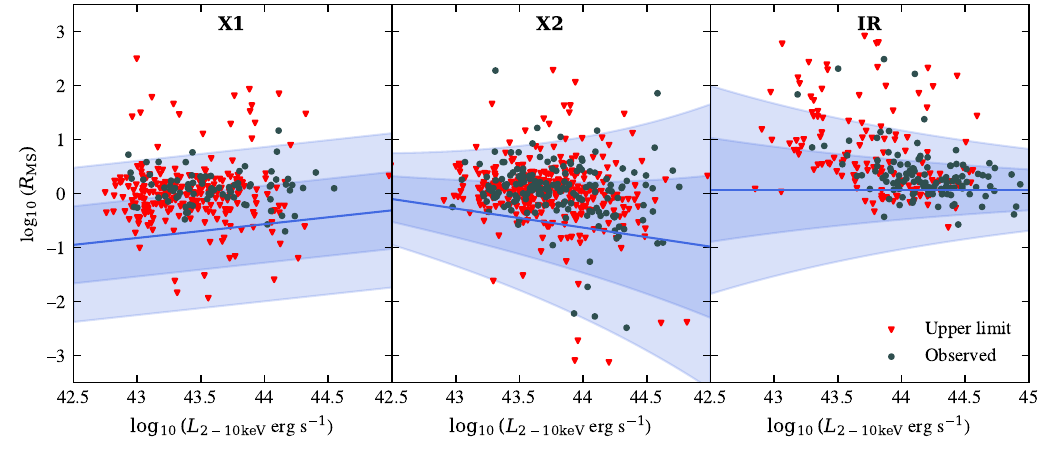}
    \caption{
        \lRMS{} against X-ray luminosity for our three samples, with upper limits denoted by red triangles. Also plotted is the Hypersampler best model (\textbf{blue line}) with 1~\&~2~$\sigma$ uncertainty contours (\textbf{blue shaded regions}). \Xg{} refers to the \lz{} X-ray-selected sample, \X{} refers to the \hz{} X-ray-selected sample, and \IR{} refers to the \hz{} IR-selected sample.
        Any slope in the Hypersampler model (\textbf{blue} line) and change in the width of the associated uncertainty contours indicates a relationship between AGN power and \RMS{}, following from the best-fitting model (described in \autoref{tab:models}) and associated parameters (shown in \autoref{tab:model_results}) for that sample as discussed in \autoref{sec:hypersampler}. 
    }\label{fig:RMS_LX}
 \end{figure*}

One way of determining whether the \lRMS{} distribution changes with AGN power is to determine the best-fitting values of $\boldsymbol{\theta}$ and evaluate whether $\theta_1$ and/or $\theta_3$ is consistent with zero (i.e., whether zero lies within their respective $1\sigma$ uncertainties). Whilst informative, a major drawback of this approach is that it does not quantify the probability that either $\theta_1$ or $\theta_3$ is equal to zero.\footnote{In the case of a continuous probability distribution, the probability of a specific value, such as zero, is infinitesimal. In such cases, only probabilities integrated over some interval are meaningful.} Instead, to quantify the posterior probability of a dependence on AGN power, we adopt the model-switching approach outlined in \G{}, and which we summarise in the next paragraph.

Within the regime described above, there are four possible models for how \lRMS{} varies with the independent variable, $X$ (X-ray luminosity): Model 0, a complex model that allows dependency in both $\mu$ and $\sigma$; Models 1 \& 2 that allow dependency in $\mu$ or $\sigma$ respectively; and 3 that does not allow any dependency. These are summarised in \autoref{tab:models}. The purpose of the Hypersampler code is to determine which model, and the corresponding most appropriate values of \Thh{}, best describes the \lRMS{} distribution and its possible dependence on \lLX{}. 

It does this by randomly switching between the four available models, with the probability of switching related to how well that model describes the data. Importantly, more flexible models are penalised during the model switching step to prevent models being preferred due to having more free parameters. The number of iterations spent in a given model is proportional to that model's posterior probability. Furthermore, whilst iterating within a given model, Hypersampler acts as a standard MCMC sampler and samples proportionally from the continuous posterior distribution of $\boldsymbol{\theta}$, which is how it simultaneously determines both the posterior probability of each model, and the best-fitting parameter values for each model.\footnote{In the case of Models 1-3 the sampler cannot deviate from a value of zero for $\theta_1$ and/or $\theta_3$, depending on the model.}

\begin{table}
    \caption{Table indicating which Hypersampler hyperparameters
    (\thh{i}) are free to vary in each of our models. Here \bf{Y} indicates the
    hyperparameter is free to vary, otherwise it is fixed at 0. When
    \thh{1}$\not=0$, then there is a \lLX{} dependence on the mean ($\mu$) of the \lRMS{}
    distribution, whereas when \thh{3}$\not=0$, then there is a
    \lLX{} dependence on width ($\sigma$) of the \lRMS{} distribution.}
    \centering
    \begin{tabular}{lllll}
    \hline
    Model & \thh{0} & \thh{1} & \thh{2} & \thh{3} \\ \hline
    0     & \bf{Y}            & \bf{Y}       & \bf{Y}            & \bf{Y}       \\
    1     & \bf{Y}            & \bf{Y}       & \bf{Y}            &              \\
    2     & \bf{Y}            &              & \bf{Y}            & \bf{Y}       \\
    3     & \bf{Y}            &              & \bf{Y}            &              \\ \hline
    \end{tabular}
    \label{tab:models}
\end{table}

\subsection{Assessing the effects of infrared selection on Hypersampler results}\label{sec:test_IR}

Again, because we use the infrared part of the SED to both identify AGN and measure their hosts' stellar masses and SFRs, we need to ensure that this does not adversely bias the results we obtain with Hypersampler. We again turn to the sample of simulated SEDs described in \autoref{sec:data:IR} to perform this assessment.

Starting with our full sample of $10^5$ simulated SEDs, we subsampled them in order to obtain two different populations: one which is curated such that the \lRMS{} distribution is independent of AGN power, and another in which the \lRMS{} distribution changes with AGN power in the same way as the \Thh{} parameters for \G{} as detailed in \autoref{tab:parameter_results}. Various other distributions were also tested (as seen in \autoref{fig:appendix_1}) and throughout Hypersampler was successful in recovering the input parameters for cases in which up to $\sim$80 per cent of the given data was upper limits. Beyond this, the uncertainties on the recovered parameters rise with increasing fractions of upper limits and/or an unbalanced distribution of upper limits (e.g. larger proportions of the lower X-ray luminosity AGN having upper limits compared to the higher X-ray luminosity AGN), as one would expect.

Next, we perform IR selection on the aforementioned two populations. This results in four subsamples:
\begin{enumerate}
    \item a non-IR-selected subsample of SEDs whose \lRMS{} distribution does not change with AGN power; 
    \item an IR-selected subsample of SEDs whose \lRMS{} distribution does not change with AGN power;
    \item a non-IR-selected subsample of SEDs whose \lRMS{} distribution changes with AGN power;
    \item an IR-selected subsample of SEDs whose \lRMS{} distribution changes with AGN power.
\end{enumerate}

We then compared the \lRMS{} distributions for the IR and non-IR-selected samples, and found no bias in the \lRMS{} distributions introduced by this selection process.


\subsection{Binning}\label{sec:binning}

\begin{figure}
    \includegraphics[width=0.95\columnwidth]{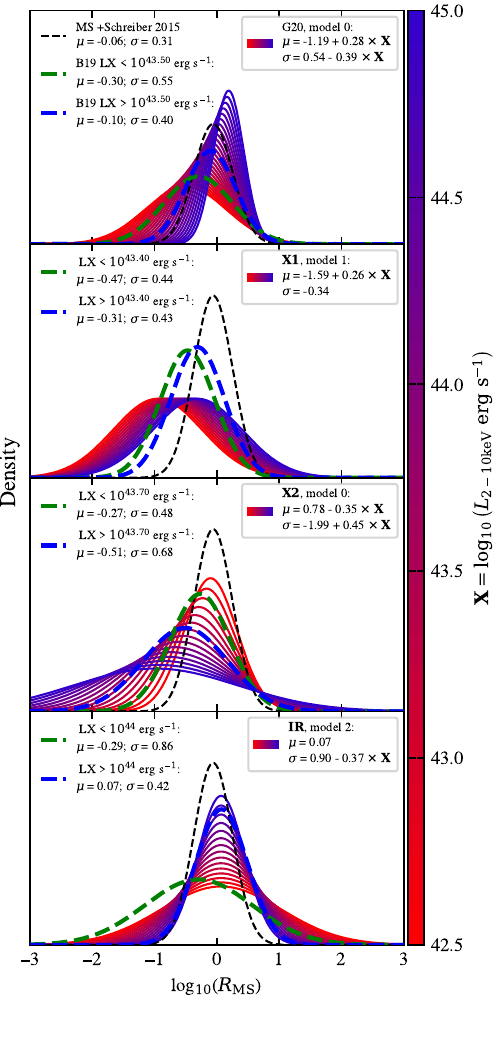}
    \caption{
        \lRMS{} distributions for \G{} (top sub-figure) and our three samples: \Xg, Model 1; \X, Model 0; \IR, Model 2, indicated by the rightmost legends. The evolution of the \lRMS{} distribution with X-ray luminosity, defined by $\bf{X}$$=\log_{10}\left(L_{\rm X}/10^{40}~{\rm ergs~s^{-1}} \right)$, in each sample is represented by the coloured (\textbf{red}-\textbf{blue}) Gaussian profiles. 
        On all figure panels the MS distribution from \protect\cite{MS} is shown in the \textbf{black dashed line}, as well as the results from a binned analysis described in \autoref{sec:binning} - similar to the analysis performed in \B{} - described in the leftmost legend.
        For \G{} (top panel), this is the results from \B, whereas for our samples these are the result of our own binned analysis.
        For the IR sample, the lower bin (\textbf{green dashed line}) is skewed by the large proportion of upper limits in this range ($\sim 70$ per cent) which cannot be accounted for in this potentially simplified Gaussian model of \lRMS{}.
        \Xg{} covers the same redshift range as \G{}, and shows similar evolution of ($\mu$) with \lLX{}, however the evolution of the $\sigma$ (spread) of the \lRMS{} distribution is not seen in the results of our analysis ($\sigma$ is constant).
        For \X{} the best fitting model for the reverses the trend from \Xg{}, whereas the distribution for \IR{} shows no evolution in $\mu$ but a similar evolution in $\sigma$ to \G{} and \Xg{}.
    }\label{fig:PDFs}
 \end{figure}

While the main statistical method used in this study is binning-free, it is useful to perform an additional binning-based analysis as both a check for our main analysis and to compare with \B. Like \B, we split our sample in half to form two equal sized bins. However, given the differences in our samples, the threshold between the two bins is set at $3.34\times 10^{43}$~erg\,s$^{-1}$, $3.70\times 10^{43}$~erg\,s$^{-1}$ and $4.00\times 10^{43}$~erg\,s$^{-1}$ for \Xg{}, \X{} and \IR{} respectively (as can be seen in the corresponding subplots in \autoref{fig:PDFs}), instead of $2.00\times 10^{43}$~erg\,s$^{-1}$ as used in \B.

We then attempt to model the \lRMS{} distributions of the samples in each bin, again assuming a log-normal distribution. The model parameters were found using a standard MCMC sampler without model switching; it is therefore limited to modelling only a single model from \autoref{tab:models}, with no way to determine which model is preferred even if multiple tests are performed using different models. 
Another disadvantage to this approach is that binning effectively assumes that all AGN in the same bin have the same X-ray luminosity, resulting in lower sensitivity to changes in the model parameters than Hypersampler - since the latter can fully account for the luminosity of each individual source.

This simplified approach of fitting static log-normal distributions is used for comparison to previous studies and as a check of Hypersampler's performance; because the sample is split according to AGN power, a strong dependence of the \RMS{} distribution on AGN power \textit{should} reveal itself across the two samples.

\begin{figure}
    \includegraphics[width=0.993\columnwidth]{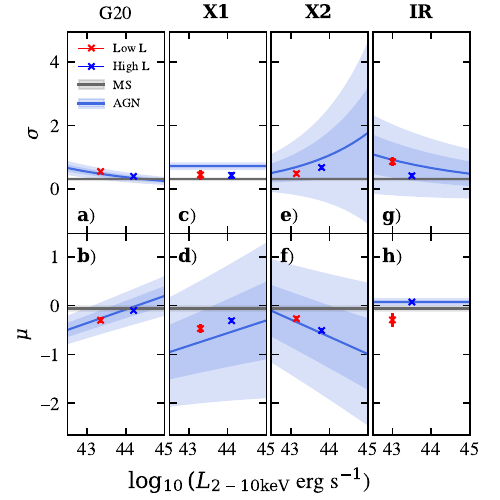}
    \caption{
        \lRMS{} distribution parameters ($\mu$, $\sigma$) from \G{} and our three samples (\Xg, \X, \IR) as they evolve with AGN power, represented here by the proxy \lLX{}. The MS $\mu$, $\sigma$ from \protect\cite{MS} are shown as a black line with a 3$\sigma$ error contour.
        The Hypersampler model parameters are shown as dark blue lines, with 1~\&~2 standard deviation error contours. 
        Results from the binned analysis described in \autoref{sec:binning} are shown as red/blue points with 1 standard deviation errors, representing the low and high AGN power bins respectively.
        For the \IR{} sample, the low AGN power bin potentially has a spurious mean ($\mu$) due to the dominance of upper limits in this bin.
    }\label{fig:mu_sig}
 \end{figure}

We therefore only use Model 3 (\thh{1}~=~\thh{2}~=~0), which assumes no dependence on AGN power, and use the difference between the best-fitting parameters for each bin to determine if there is a dependence (as can be seen in the binned results in \autoref{fig:PDFs} \& \autoref{fig:mu_sig}). 

This binning method does, however, account for upper limits on \RMS{} by taking the same approach (using the CDF in addition to PDF) as described above. 
Tests based on simulated data showed that - like Hypersampler - this method struggles to return the true underlying parameters of the distribution when there is a very high proportion of upper limits (>80 per cent).

\section{Results}\label{sec:results}

In this section, we outline the results that we obtained from running Hypersampler on both our X-ray-selected and IR-selected samples of AGN, as well those obtained from our binned samples (see \autoref{sec:binning}). For all samples, the output of Hypersampler (a chain of model identification numbers and \Thh{} values) is thinned to ensure adjacent samples are independent. The parameter values obtained by our Hypersampler analysis are summarised in \autoref{tab:model_results} and are described below. 

\begin{table}
    \caption{Proportions of MCMC samples spent in each model (described in \autoref{sec:hypersampler}) for \G{} and across our three samples (\Xg{}, \X{}, \IR{}). The listed value represents the probability of each model, and sum to 1 for each sample of AGN, drawn from the proportion of samples the MCMC component of Hypersampler spends in each model. \Xg{} refers to the \lz{} X-ray-selected sample, \X{} refers to the \hz{} X-ray-selected sample, and \IR{} refers to the \hz{} IR selected sample.
    For models with \thh{1}~=~0, $\mu$ is therefore independent of \lLX{} and constant; similarly, for models with \thh{3}~=~0, $\sigma$ is constant.}
    \centering
    \begin{tabular}{lcccc}
    \hline

    Sample & model 0 & \multicolumn{1}{p{1cm}}{\centering model 1: \\ $\theta_{1}=0$} & \multicolumn{1}{p{1cm}}{\centering model 2: \\ $\theta_{3}=0$} & \multicolumn{1}{p{2cm}}{\centering model 3: \\ $\theta_{1}=\theta_{3}=0$} \\ \hline

    \G{}            & 0.933               & 0.005             & 0.002          & 0.061              \\
    \textbf{X1}     & 0.071          & 0.434        & 0.112     & 0.383         \\
    \textbf{X2}     & 0.991          & 0.004        & 0.073     & 0.011         \\
    \textbf{IR}     & 0.128          & 0.005        & 0.849     & 0.018         \\ \hline

    \end{tabular}
    \label{tab:model_results}
\end{table}

\subsection[\Xg{}: X-ray-selected, 0.8 < z < 1.2 sample]{\Xg{}: X-ray-selected, \lz{} sample}\label{sec:lowz}

This sample is similar by design to that used in \B{} \& \G{} in that it is comprised of X-ray-selected AGN within the redshift range of \lz{}. However, in the time since \G{} performed their analysis of X-ray-selected AGN, both new photometric catalogues and an updated version of \C{} have been released, making it worthwhile to re-visit this region of parameter space (X-ray-selected, \lz{}). 

In panels (\textbf{a}) and (\textbf{b}) of \autoref{fig:mu_sig} we present the results of \G{} in terms of how the parameters ($\mu$ \& $\sigma$) of the \lRMS{} distribution change with X-ray luminosity for their X-ray-selected, \lz{} sample. Their analysis shows that both $\mu$ and $\sigma$ change with X-ray luminosity. By contrast, our re-analysis of AGN in this same parameter space (X-ray-selected, \lz{}) reveals that our Model 1 - in which $\mu$ changes with X-ray luminosity, but $\sigma$ does not - has the highest posterior probability of all four models at 0.434 (see \autoref{tab:model_results}). This is not significantly higher than that of model 3 (posterior probability: 0.383, $\mu$=-0.690, $\sigma$=-0.336), which has no dependency on \lLX{}. This means that while our model 0 shows the same degree of dependence of $\mu$ on \lLX{} as \G{}, it is at a far lower level of significance as reported in that study.

The increasing $\mu$ with increasing X-ray luminosity can be seen in panel (\textbf{d}) of \autoref{fig:mu_sig}, though comparing it to panel (\textbf{b}) reveals that this relationship has greater uncertainties (and is thus less significant) than that found by \G{}, which aligns with our greater likelihood of model 3 being favourable compared to \G{}.

The results from our analysis of the binned X-ray-selected, \lz{} sample (see \autoref{sec:binning}) supports the results from Hypersampler, although in both bins $\mu$ and $\sigma$ tend to be closer to 0 than Hypersampler predicts. These are shown as points in \autoref{fig:mu_sig} and, again, show evidence of a mild increase in $\mu$ with increasing X-ray luminosity, but no statistically significant change in $\sigma$.

To help visualise the consequence of the the above findings on the \RMS{} distributions of our samples, we show in \autoref{fig:PDFs} how those distributions themselves change as a function of X-ray luminosity. The narrowing and shifting of the distributions implied by the \G{} results are clear in the top subplot (\G{}), while the second subplot (\Xg) shows only a shifting of the distribution centre as implied by our findings.

\begin{table}
    \caption{Results from \G{} and our best-fitting Hypersampler results for \Thh{} across our three samples. \Xg{} refers to the X-ray-selected sample between redshifts of \lz{}, \X{} and \IR{} refer to AGN selected between redshifts of \hz{} for X-ray and IR selected AGN respectively. Uncertainties quoted here are 1$\sigma$, and symmetric in logspace as the hyperparameters (\Thh{}) are defined by \autoref{eq:mu_sigma} and our assumption of a lognormal \RMS{} distribution.}
    For both \Xg{} \& \X{}, the key hyperparameters (\thh{1} \& \thh{3}) are not consistent with 0 at a 3$\sigma$ significance, indicating that the samples show evidence for a dependence of \lRMS{} on \lLX{} in both $\mu$ \& $\sigma$. For \IR{}, \thh{1}$=0$, meaning no evidence of a dependence in the $\mu$ parameter, but \thh{3}$\not=0$ meaning that the $\sigma$ parameter does. This can be seen in \autoref{fig:PDFs} \& \autoref{fig:mu_sig}.
    \centering
    \begin{tabular}{lcccc}
    \hline
    Sample              & $\theta_{0}$                  & $\theta_{1}$              & $\theta_{2}$              & $\theta_{3}$              \\ \hline
    \citetalias{G20}    & -1.19$\pm0.12$           & 0.28$\pm0.03$        & 0.54$\pm0.13$        & -0.39$\pm0.04$       \\
    \textbf{X1}         & -1.59$\pm0.45$      & 0.26$\pm0.13$   & -0.34$\pm0.08$  & 0                    \\
    \textbf{X2}         & 0.78$\pm0.43$       & -0.35$\pm0.12$  & -2.00$\pm0.48$   & 0.45$\pm0.12$   \\
    \textbf{IR}         & 0.07$\pm0.04$       & 0                    & 0.90$\pm0.45$   & -0.37$\pm0.11$  \\ \hline
    \end{tabular}
    \label{tab:parameter_results}
\end{table}

\subsection[X-ray \& IR samples (X2 \& IR)]{X-ray \& IR samples (\X{} \& \IR{})}\label{sec:highz}
We now move on to the analysis of our higher redshift (\hz{}) X-ray and IR-selected samples of AGN. As with the \X{} sample discussed in the previous section, we also split these samples into two bins for a complimentary analysis - this time at X-ray luminosities of $3.34\times 10^{43}$~erg\,s$^{-1}$, $3.70\times 10^{43}$~erg\,s$^{-1}$ and $4.00\times 10^{43}$~erg\,s$^{-1}$ for \Xg{}, \X{} and \IR{} respectively to maintain roughly equal numbers in each bin within each sample.\footnote{As described in \autoref{sec:hypersampler}, in the case of the IR-selected sample we convert the 12~$\mu m$ luminosity to X-ray luminosity.}
\subsubsection{\X{}: X-ray-selected \hz{} sample}\label{sec:results:X_highz}

We show in panels (\textbf{e}) and (\textbf{f}) of \autoref{fig:mu_sig} and the \X{} panel of \autoref{fig:PDFs} the outcome of our analyses of the \X{} sample. 
These plots show that, in contrast to the results from \G{} and our own lower redshift sample sample, a negative trend in the $\mu$ parameter with \lLX{} is seen - as shown by the \thh{1} hyperparameter for sample \X{} in \autoref{tab:parameter_results}, -0.353$\pm$0.115 ($>3\sigma$ away from 0). As the \thh{3} value is positive, this results in the \X{} panel in \autoref{fig:PDFs} being visually similar to \G{}'s result in the top panel, but inverted - with lower power AGN tending to occur in galaxies with higher \lRMS{} values.

\autoref{tab:model_results} shows that model 1 is far more preferred over any other model, with 99.1 per cent of samples - and therefore an estimated posterior probability of model 1 being accurate of 0.991. This gives a Bayes factor (a ratio of probabilities) of $\sim$13.6 over the next most likely model (model 2) and $\sim$90 over model 3.

Both \autoref{fig:PDFs} \& \autoref{fig:mu_sig} show the results of a binned analysis similar to \B, with an X-ray luminosity split between the two bins of $10^{43.7}$; the parameters of these binned samples agree strongly with Hypersampler.

\subsubsection{\IR{}: IR-selected \hz{} sample}\label{sec:results:IR}

The IR-selected AGN sample comes with a set of additional complications. 
One complication is these AGN are selected based on the shape of their SEDs in the IR part of the spectrum, yet the flux at these wavelengths also contribute to our measurement of the stellar mass and SFRs of the host galaxy. As such, we need to be careful that this is not introducing biases into our results. This is covered in detail in \autoref{sec:test_IR}, with the outcome being that we are confident that IR selection is not biasing the results described below.

A further complication associated with the IR-selected sample is the larger proportion of \RMS{} upper limits (vs detections) at the lower end of the X-Ray luminosity range, as seen in \autoref{fig:RMS_LX}. Since Hypersampler considers the full dataset (as opposed to subsets, i.e. bins) at once, it is less affected than the binned analysis by regions of parameter space that are dominated by upper limits. By contrast, the analysis of our lower luminosity ($L_{2-10\text{keV}} < 10^{44}$~erg\,s$^{-1}$) bin is strongly affected by the large proportion (70 per cent) of upper limits in that bin.  
 
When we run Hypersampler on the full IR-selected sample, it strongly prefers Model 2 (only $\sigma$ depend on AGN power) over all other models as seen in \autoref{tab:model_results}. 

This suggests that there exists a relationship between AGN power and the \RMS{} distribution, but only in that higher power AGN are more likely to occur in galaxies with \lRMS{}$\sim0$, whereas lower power AGN can occur in a much wider range. It is worth noting the increased proportion of upper limits in the lower \lLX{} end of this sample, which would naturally lead to an increased level of uncertainty. In this sample, lower power AGN tend to exist in these galaxies with less dominant SFR (in the photometric bands used to estimate SFR) and thus have upper limits on their SFRs. This may be due to lower common fuel reserves (gas, dust) in these galaxies leading to a reduction in both AGN activity and SFR.

\section{Discussion}\label{sec:discussion}

Following a thorough validation and testing process of selection and analysis techniques, we have analysed three samples of active galactic nuclei in order to uncover subtle relationships between AGN power and host galaxy SFRs. Our ultimate aim is to shed light on the complex interplay between AGN and their host galaxies.

Using updated tools and photometric data, we have repeated the analyses performed by \B{} \& \G{}. Their sample and results listed in \autoref{tab:parameter_results} and shown in \autoref{fig:PDFs} \& \autoref{fig:mu_sig} labelled as \G{} and/or \B, whereas our complimentary sample is identified as \Xg. These earlier studies found that both the centre and spread of the \lRMS{} distribution varied with AGN power - shown by all \Thh{} values being non-zero (summarised in \autoref{tab:parameter_results}) and the varying centre and width of the Gaussian profiles plotted in the top subplot in \autoref{fig:PDFs}. 

Our results for the \Xg{} sample differ from those of \G{} in that our most favoured model is one in which the centre ($\mu$) of the \lRMS{} distribution varies with AGN power but not the spread ($\sigma$). This can be seen by \thh{3}$=0$ in \autoref{tab:parameter_results}, and the constant width of the Gaussian profiles in the \Xg{} subplot (second from top) in \autoref{fig:PDFs}.

While this is a different result to \B{} \& \G{}, the core result - that there is a SFR dependancy on AGN power - is retained, albeit at a much lower level of significance. This lower level of significance can be seen both in terms of the similarity between the proportions of samples Hypersampler spent in each model (0.434 for model 1 vs. 0.383 for model 3; i.e., only a very mild preference for the \lLX{}-dependent model 1), and in terms of the 1$\sigma$ uncertainties on the \Thh{} values seen in \autoref{tab:parameter_results}. By contrast, the equivalent sampling proportions reported by \G{} are 0.93 and 0.06 in favour of their \lLX{}-dependent model. 

Since both Hypersampler and the input data -- specifically the optical-to-near-IR photometric data -- and \C{} have changed between this study and \G{}, it is challenging to ascertain the root cause of the fall in the significance of the \lLX{}-dependent model between this and that previous study. However, some insight can be obtained by comparing the results of our binning analysis to that of \B, who used the same input data as \G{}. The results of our binning analysis are consistent with our Hypersampler results, at least in terms of the general trends or lack thereof (seen in \autoref{fig:mu_sig} and \autoref{fig:PDFs}). Since our binning analysis is independent of Hypersampler, this suggests that the cause of the drop in significance between this study and \G{} lies in the differences in the input data. But given the size of the photometric catalogues and the large number of adjustable \C{} parameters, we consider it beyond the scope of this study to investigate the specific root cause of the reduced significance of the \lLX{}-dependent model between \G{} and this study.

Whilst the \Xg{} sample was selected to directly compare to \G{}, with the overall aim to compare the differences between X-Ray and IR-selected AGN power and \lRMS, there were too few IR-selected AGN within the original redshift range (\lz) to perform a meaningful analysis. To remedy this, we created two new samples -- \IR{} \& \X{} -- the first being an IR-selected sample spanning a redshift range of \hz, and the second an X-ray-selected sample in the same redshift range to act as a comparison sample (see \autoref{sec:samples}).

Our results from running Hypersampler on the \X{} sample reveal strong evidence for a negative correlation between AGN power and \lRMS; i.e., more powerful AGN tend to live in galaxies with lower SFR relative to the galaxy main sequence. This is in direct contrast to the results reported above for our \Xg{} sample which shows the opposite, at low significance, or no evidence for a significant relationship between \lRMS and \lLX{}. It also follows the opposite trend to that reported by \G{} for their lower-redshift X-ray-selected sample.

The \IR{} sample presented numerous challenges; the primary barrier to a robust and validated result (i.e. Hypersampler's binning-free approach agreeing with a binned analysis) is the large proportion of upper limits (>70 per cent) on the SFR in the low AGN power range, visible in \autoref{fig:RMS_LX}. This could lead to the binned analysis underestimating $\mu$ and over-estimating $\sigma$, even if our analysis of the binned data can, too, account for upper limits. \autoref{fig:PDFs} shows the \lRMS{} distribution of this lower luminosity bin with a $\sigma$ significantly higher than the PDF constructed from the hyperparameters for \lLX{}$=42.5$, which is the lower end of the luminosity range of our \IR{} sample. Through extensive testing during which synthetic data that included various proportions of upper limits, Hypersampler demonstrated that it can largely account for an unbalanced distribution of upper limits within a sample -- up to $\sim80$ per cent upper limits regardless of where they are placed, with a commensurate increase in uncertainty as can be seen in \autoref{fig:appendix_1}.\footnote{Here, ``unbalanced'' refers to there being different proportions of upper limits at lower \lLX{} values compared to higher \lLX{} values.} Testing showed that Hypersampler was able to return hyperparameters that are consistent with the values used to generate the synthetic samples, although with increasing uncertainties as the proportion of upper limits was increased, as should be expected.

In the case of the \IR{} sample, \thh{2} is $<2\sigma$ from 0, indicating that the most preferred model (2) is not significantly different from model 3. Having said that, the sampling proportions show that model 2 is preferred at the level of $84.9$ per cent, with a Bayes factor of 47 over \lLX{}-independent model 3. Too few samples in the final chain were spent in model 3 to reliably estimate \Thh{} values, preventing further comparison.

The two redshift ranges covered by our samples include roughly the same amount of cosmic time: \lz{} spans from roughly 6.8~Gyr to 5.2~Gyr after the Big Bang and \hz{} spans roughly 5.9~Gyr to 3.7~Gyr after the Big Bang. These ranges overlap but \hz{} includes much more of the early Universe, and the continuous relationships uncovered by Hypersampler should account for cross-contamination between the samples. This may indicate that in the early Universe, more powerful IR-selected AGN reside in galaxies that span a narrow \lRMS{} range centered on the main sequence, whereas lower power AGN are found in a more diverse range of galaxies. By contrast, the higher luminosity X-ray-selected AGN reside in galaxies with a broader range of \lRMS{} values than their lower power counterparts.

It has been speculated that heavily obscured AGN represent an earlier phase in a galaxy's evolution. The fact that our \IR{} sample, which likely includes such heavily obscured AGN (see \citealt{review_andonie}), behaves differently to the X-ray-selected sample at the same redshifts could be seen as supporting this hypothesis. IR-selected AGN may occur in gas-rich galaxies, similar to the galaxy main sequence, with relatively high SFRs; whereas higher power X-ray-selected AGN tend to occur in less rapidly star forming galaxies, perhaps indicating that X-ray AGN contribute to the reduction of SFR by disrupting the galaxies reserves of cold, dense gas - at least when they are active.

In the more recent universe, at lower redshift, we do not see this trend amongst X-ray galaxies - with either a positive relationship or no relationship exists - meaning that either powerful AGN live in galaxies with large supplies of cold, dense gas for fuel that \textit{also} fuels increased SFRs, or there is no trend and there is little interplay between nuclear activity and star formation in this redshift regime.

As some of the IR-selected AGN \textit{do} have X-ray detections, it may be that these unobscured, or less strongly obscured, AGN lie in the lower \RMS{} region of the sample's parameter space. Conversely, the heavily obscured AGN discussed in \cite{review_andonie} may make up the higher SFR regions. However, because we have not considered X-ray detections as a measure of obscuration amongst our \IR{} sample, more detailed investigation into the link between obscuration and star formation is warranted.

Galaxies and their SMBHs evolve over time, consuming the raw gas and dust to form stars and power AGN; mergers and secular processes can accelerate this, leading to the extreme variation in galaxy properties in a large, many-dimensional parameter space. However, due to the random nature of the effect these processes have, the range of astrophysical properties exhibited by galaxies can be understood on statistical terms using large samples. The trends we see in our samples are therefore indicative of common trends that affect all galaxies during their evolution. Our conclusions are therefore also rather broad; we can be sure that a number of factors have affected the evolution of galaxies, but not of the relative contributions of each factor. For example, a high gas/dust content of a galaxy can lead to higher SFR and AGN growth but the complex interplay over the timescale of human observation is unclear - here we see over the course of Gyr there is a relationship between AGN power (which is tied to AGN growth) and SFR.

\section{Conclusions}\label{sec:conclusions}

Through our analysis and testing of our methods, we have found the following key findings:

\begin{itemize}

    \item IR selection of AGN is robust across redshift and other galactic properties;
    \item The properties of the \lRMS{} distribution of X-ray-selected AGN at \lz{} likely differ from X-ray-selected AGN at \hz{}, with the \lRMS{} distribution of the former showing no statistically significant trend with AGN power while the latter show a statistically significant negative trend in $\mu$ (i.e., the location of the peak of the distribution);
    \item The \lRMS{} distribution of IR-selected AGN within the \hz{} redshift range differs from X-ray-selected AGN within the same redshift range in that the IR-selected AGN show no statistically significant relationship between $\mu$ and AGN power, but one does exist for $\sigma$ (i.e., the width of the distribution).

\end{itemize}

The relationship between AGN power and the host galaxy sSFR is not clear, varying with our different samples. For example, there is weak but likely not statistically significant sign of a positive relationship - i.e. no sign of more powerful AGN causing simultaneous reductions in host SFR - at lower redshifts, but this is observed in X-ray-selected AGN at higher redshifts. In contrast, IR-selected AGN do not show this trend. As IR-selection is believed to be more robust against obscuration, it may be that the IR-selected AGN exist in more gas-rich galaxies that can support both AGN and star formation, whereas the higher-power X-ray-selected AGN may exist in less gas rich galaxies, and perhaps contributed to the fact they are less gas rich. However, due to the different timescales of AGN activity and star formation, it may be that any quenching effect - or vice versa - is not visible here as the effects of the active AGN are not typically seen while the AGN is active, and selected as such.

With the exception of the high redshift, IR-selected \IR{} sample, we see that the \lRMS{} distributions of the AGN are notably different from the MS, and with the exception of the low redshift, X-ray-selected \Xg{} sample, likely have some dependency on AGN power.

\section*{Acknowledgements}

We thank the anonymous referee for their comments. We are also grateful to Paul Kerry for his assistance with technical aspects of this research.

\section*{Data Availability}

All photometric, spectroscopic and extensive properties (e.g., redshift) data used in this work is drawn from archival and openly available data that is publicly available as listed in \autoref{sec:data}. Simulated data is created using \C{}, the use of which is detailed in \autoref{sec:mass} using the parameter ranges listed in \autoref{tab:cigale}.


\bibliographystyle{mnras}
\bibliography{bibl} 
\newpage
\appendix
\onecolumn

\begin{table}
    \caption{
    During analysis, we use the following parameters to construct our grid for SED fitting using \C{}. This is not an exhaustive list, as we only list the parameters that we alter from their default values.
    }
    \label{tab:cigale_analysis}
    \centering
    \begin{tabular}{ll}
    \hline
    Module \& parameters &  \\ \hline

    \textbf{Star formation history} & \cite{cigale}                     \\
    $\tau_{main}$                       & $100, 1000, 3000, 1\times 10^{10}$ Myr   \\
    age$_{main}$                        & $100, 250, 500, 1000, 3000, 4500, 6000, 7500, 9000, 12000$ Myr             \\
    $\tau_{burst}$                      & $50$ Myr                \\
    age$_{burst}$                       & $20$ Myr                \\
    $f_{burst}$                         & $0.0, 0.05, 0.10, 0.20$                \\ \hline

    \textbf{Stellar population synthesis} & \cite{bruzual2003stellar}                    \\
    IMF                                 & 1 (\citealt{chabrier2003galactic})             \\ 
    Metallicity                         & 0.02 \\ 
    Separation age                      & $10$ Myr            \\ \hline

    \textbf{Dust attenuation} & \cite{calzetti2001dust}                    \\
    E(B-V)* young stellar population    & $0.01, 0.05, 0.1, 0.2, 0.3, 0.4, 0.5, 0.6, 0.7, 0.8, 0.9, 1.0, 1.1, 1.2, 1.4$              \\
    E(B-V)* old stellar population      & $0.44$             \\\hline

    \textbf{Dust emission} & \cite{dale2014}                    \\
    $\alpha$ slope                      & 1.5, 2.5 \\ \hline 

    \textbf{AGN} & \cite{fritz2006}                    \\
    Torus max-min radius ratio          & $60$   \\
    $\tau$ - 9.7$\mu$m optical depth    & $1.0, 6.0$   \\
    $\beta$                             & $-0.5$   \\
    $\gamma$                            & $4.0$   \\
    opening angle                       & $100^{\circ}$               \\  
    $\psi$ (viewing angle)              & $0.001, 50.100, 89.990$   \\
    \fracAGN{} & $0.01, 0.1, 0.2, 0.3, 0.4, 0.5, 0.6, 0.7, 0.8, 0.9, 0.99$                  \\ \hline

    \end{tabular}
    
\end{table}

\section{Hypersampler testing}\label{sec:appendix_HS}

 \begin{figure}
    \includegraphics[width=0.993\columnwidth]{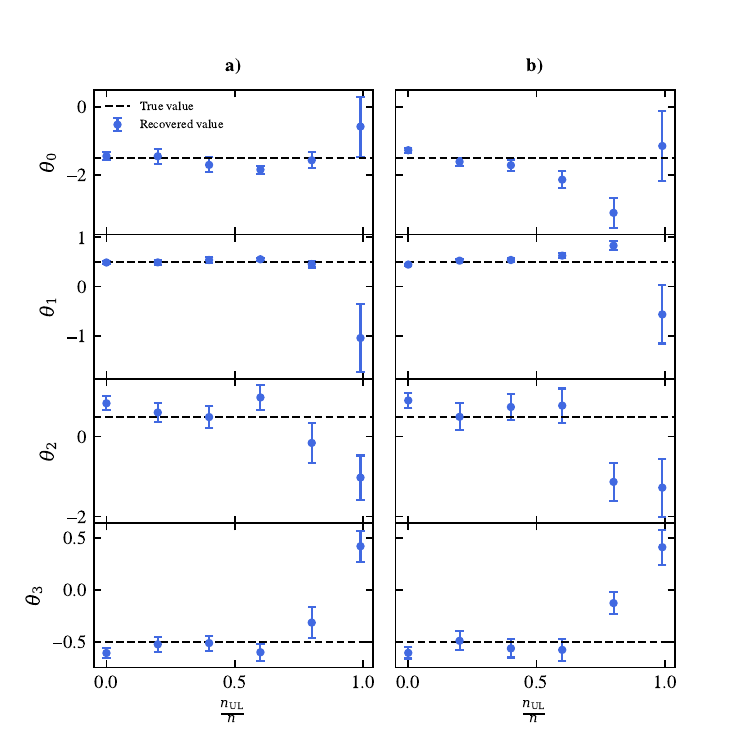}
    \caption{
        Hypersampler testing on simulated datasets with varying proportions of upper limits ($\frac{n_{\text{UL}}}{n}$, where $n$ represents the full dataset and $n_\text{UL}$ represents the number of upper limits.). The simulated dataset had all $\theta$ hyperparameters set to non-zero values (i.e., equivalent to our Model 0). The left-hand column refers to datasets where upper limits are uniformly distributed at random throughout the dataset, whereas the right-hand column refers to datasets with upper limits preferentially placed at lower X-ray luminosities. The error bars represent 1$\sigma$ uncertainties. Up to upper limit proportions of around 80 per cent, Hypersampler is capable of reliably recovering the values of all four hyperparameters. For reference, the proportion of upper limits in our samples are \Xg: 74 per cent, \X: 64 per cent and \IR: 52 per cent. For this analysis, Hypersampler had model switching disabled and had a limited run of 2000 samples.
    }
    \label{fig:appendix_1}
 \end{figure}

\label{lastpage}
\end{document}